\documentclass{JHEP3}
\usepackage{amssymb}
\usepackage{amsfonts}
\usepackage{amsmath}
\usepackage{revsymb}
\usepackage{cite}
\usepackage{graphicx}

\newcommand{\be}{\begin{equation}}
\newcommand{\ee}{\end{equation}}

\newcommand{\bse}{\begin{subequations}}
\newcommand{\ese}{\end{subequations}}
\newcommand{\bea}{\begin{eqnarray}}
\newcommand{\eea}{\end{eqnarray}}
\newcommand{\ba}{\begin{array}}
\newcommand{\ea}{\end{array}}

\def\P{Poincar\'e }
\def \th {\theta^{\mu\nu}}

\title{{Gauge field theories with covariant star-product}}
\author{M. Chaichian$^1$, A. Tureanu$^1$ and G. Zet$^2$\\

$^1$Department of Physics, University of Helsinki and Helsinki
Institute of Physics, P.O.Box 64, FIN-00014 Helsinki, Finland\\
$^2$Department of Physics, "Gh. Asachi" Technical University,\\Bd.
D. Mangeron 67, 700050 Iasi, Romania}

\abstract {A noncommutative gauge theory is developed using a
covariant star-product between differential forms defined on a
symplectic manifold, considered as the space-time. It is proven that
the field strength two-form is gauge covariant and satisfies a
deformed Bianchi identity. The noncommutative Yang-Mills action is
defined using a gauge covariant metric on the space-time and its
gauge invariance is proven up to the second order in the
noncommutativity parameter.}

\preprint{HIP-2009-9/TH}

\dedicated{Dedicated to Ioan Gottlieb on the occasion\\ of his 80th
birthday anniversary.}

\begin{document}

\section{Introduction}

Noncommutative gauge theories have been intensively studied recently
(see \cite{4}-\cite{13} for an incomplete list of references). They
are defined on noncommutative space-times whose coordinates satisfy
the property
\be [x^\mu,x^\nu]=i\theta^{\mu\nu}, \label{1.1}\ee
where in the canonical case $\theta^{\mu\nu}$ is an antisymmetric
constant matrix of dimension length squared. The gauge fields on
such a space-time are considered functions of the noncommutative
coordinate operators. Through Weyl-Moyal correspondence quantization
the noncommutative algebra of operators generated by \eqref{1.1} can
be represented on the algebra of ordinary functions on classical
space-time by using the noncommutative Moyal $\star$-product. The
gauge theories defined by the $\star$-action of the gauge algebra
generators obey a no-go theorem, which strongly restricts the model
building \cite{NCSM, NCSM-Wess}.

As $\theta^{\mu\nu}$ is constant, the Lorentz invariance of
\eqref{1.1} and, consequently, of the corresponding field theory
built on such a space-time, breaks down. Nevertheless,
noncommutative field theories in general, and gauge theories in
particular formulated with Moyal $\star$-product, are invariant
under the twisted Poincar\'e symmetry \cite{CKNT,32}. In the case of
twist-deformations, the generators of the twisted Hopf algebra act
as usual on individual fields (see \cite{CKTZZ,CNST} for a detailed
account of the action of twisted \P algebra and the meaning of
twisted \P invariance), which leads to the natural assumption that
the no-go theorem for noncommutative gauge theories could be
circumvented if the action of the gauge generators were to be
expressed by a twist. However, it was proven \cite{13} that the
concept of twist symmetry, originally obtained for the
noncommutative space-time \eqref{1.1}, cannot be extended to include
internal gauge symmetry. In other words, it is not possible to
obtain a gauge covariant twist if the property \eqref{1.1} is
adopted with $\th$ constant.

    The same result appears to be valid also in the case of noncommutative gauge theory of gravitation.
    In Ref. \cite{12} it has been shown that the twisted Poincar\'e symmetry cannot be gauged by generalizing
    the Abelian twist to a covariant non-Abelian twist, nor by introducing a more general covariant
    twist element defined with  $\th$ constant. Other methods used to formulate a noncommutative theory of
    gravitation \cite{23,21, 10,24,22,33} suffer from the same difficulty or from the restrictions of the no-go theorem.

    As the introduction of a gauge covariant twist, defined with $\th$  constant, breaks the associativity of the
    algebra of functions on noncommutative space-time, both in the internal and external gauge symmetry cases, we
    may have to consider space-time geometries that are also non-associative, not only noncommutative. Indeed, there
    exist in the literature works on constructing non-associative theories with some desired properties (see, e.g.,
    \cite{11, 27, 28, 29, 30} and references therein). However, non-associativity introduces many difficulties in
    formulating gauge models and they are practically non-attractive.

    One possible way to define a covariant star-product satisfying the associativity property is to consider models
    of noncommutativity with $\th$ depending on coordinates. In Ref. \cite{1} a new covariant star-product between differential
    forms has been defined. For ordinary functions, which are differential forms of order zero, this product reduces to that
    given by Kontsevich \cite{5} (see also \cite{6} for results up to the fourth order in $\theta$   and \cite{8} for a path integral approach). The property
    of associativity of the new covariant star-product has been explicitly verified up to the second order in the noncommutativity parameter in Ref. \cite{1}.

    In this paper we extend the definition given in Ref. \cite{1} to the case of Lie algebra valued differential forms, with the ultimate aim of constructing noncommutative
    gauge field theories. Thus we obtain a graded
    Lie algebra valued Poisson algebra where the star-bracket operation can be both commutator and anti-commutator, depending on the grades of
    the two forms and the order in $\theta$ of the considered term in the star-product. The space-time is supposed to be a symplectic manifold on
    which a Poisson bracket is defined.

In Section 2 we give the definition of the star-product between two
arbitrary Lie algebra valued differential forms and some of their
properties. Then, the star-bracket   between such differential forms
is introduced and some examples are given. Section 3 is devoted to
the noncommutative gauge theory formulated with the new gauge
covariant star-product. The noncommutative Lie algebra valued gauge
potential and the field strength two-form are defined and their
gauge transformation laws are established. It is shown that the
field strength is gauge covariant and satisfies a deformed Bianchi
identity.   In Section 4 a gauge covariant noncommutative metric on
the space-time manifold is introduced and the action for the gauge
fields is written using the star-product. It is proven that this
action is gauge invariant to the second order in $\theta$. Section 5
is dedicated to the discussion of the results and to an
interpretation of noncommutative gauge theory formulated by using
the star-product between Lie algebra valued differential forms on
symplectic manifolds.

\section{ Definition of star-product}

We consider a NC space-time $M$ endowed with the coordinates
$x^\mu$, $\mu=0,1,2,3$, satisfying the commutation relation
\be[x^\mu,x^\nu]=i\theta^{\mu\nu}(x),\label{2.1}\ee
where $\theta^{\mu\nu}(x)=-\theta^{\nu\mu}(x)$  is a Poisson
bivector \cite{1}. Define then the Poisson bracket between two
functions $f(x)$ and $g(x)$ by
\be\{f,g\}=\th\,\partial_\mu f\,\partial_\nu g\,.\label{2.2}\ee
In order for the Poisson bracket to satisfy the Jacobi identity, the
bivector $\theta^{\mu\nu}(x)$  must obey the condition
\be
\theta^{\mu\rho}\partial_\rho\theta^{\nu\sigma}+\theta^{\nu\rho}\partial_\rho\theta^{\sigma\mu}+\theta^{\sigma\rho}\partial_\rho\theta^{\mu\nu}=0\,.
\label{2.3}\ee
If a Poisson bracket is defined on $M$, then $M$ is called a Poisson
manifold (see \cite{7} for mathematical details).
    Suppose now that the bivector  $\theta^{\mu\nu}(x)$ has an inverse  $\omega_{\mu\nu}(x)$, i.e.
\be \theta^{\mu\rho}\omega_{\rho\nu}=\delta^\mu_\nu\,.\label{2.4}\ee
If $\omega=\frac{1}{2}\omega_{\mu\nu} dx^\mu\wedge dx^\nu$  is
nondegenerate ($\det \omega_{\mu\nu}\neq 0$) and closed
($d\,\omega=0$), then it is called a symplectic two-form and $M$ --
a symplectic manifold. From now on we denote the exterior product of
two forms $\alpha$ and $\beta$ simply by $\alpha\beta$ and
understand that it means $\alpha\wedge\beta$. It can be verified
that the condition $d\,\omega=0$ is equivalent to the equation
\eqref{2.3} \cite{1, 2}. In this paper we shall consider only the
case when $M$ is symplectic.

    Because the gauge theories involve Lie algebra valued differential forms
    such as $A=A_\mu^a\, T_a\, dx^\mu=A_\mu\,dx^\mu$, $A_\mu=A_\mu^a(x)\, T_a$, where $T_a$  are the infinitesimal generators of a symmetry Lie group $G$,
    we need to generalize the definition of the Poisson bracket to differential
    forms and define then an associative star-product for such cases. These
    issues were solved in Ref. \cite{1, 2, 3} and here we just recall the definitions
    and properties to fix the idea. However, we generalize these results to the case of
    Lie algebra valued forms. This means that the Poisson algebra becomes a graded Lie algebra valued
    one. Therefore, the commutator of differential forms can be a commutator or
    an anti-commutator, depending on their degrees.

    Assuming that $\th(x)$  is invertible, we can always write the Poisson bracket  $\{x,dx\}$ in the form \cite{1,2}
\be
\{x^\mu,dx^\nu\}=-\theta^{\mu\sigma}\Gamma^\nu_{\sigma\rho}dx^\rho\,,\label{2.5}\ee
where $\Gamma^\nu_{\sigma\rho}$ are some functions of $x$
transforming like a connection under general coordinate
transformations. As  $\Gamma^\nu_{\sigma\rho}$ is generally not
symmetric, on can use the one-forms connection
\be\widetilde \Gamma^\mu_{\nu}=\Gamma^\mu_{\nu\rho}dx^\rho, \ \ \ \
\Gamma^\mu_{\nu}=dx^\rho\Gamma^\mu_{\rho\nu}\label{2.6}\ee
to define two kinds of covariant derivatives, $\widetilde\nabla$ and
$\nabla$, respectively. For example, if  $\alpha=\alpha_\nu\,
dx^\nu$ is a one-form, then
\be
\widetilde\nabla_\mu\alpha=(\partial_\mu\alpha_\nu-\Gamma^\rho_{\mu\nu}\alpha_\rho)\,dx^\nu\label{2.7}\ee
and analogously for  $\nabla_\mu\alpha$. Given  $\theta$ and
$\Gamma$, all Poisson brackets are determined \cite{2}. Now, if
$\alpha$ and $\beta$ are two arbitrary differential forms, then
their Poisson bracket is given by \cite{1}
\be\{\alpha,\beta\}=\th\nabla_\mu\alpha\nabla_\nu\beta-\widetilde
R^{\mu\nu}(i_\mu\alpha)(i_\nu\beta)\,,\label{2.8}\ee
where
\be \widetilde R^{\mu\nu}=\frac{1}{2}\widetilde
R^{\mu\nu}_{\rho\sigma}\,dx^\rho dx^\sigma\label{2.9}\ee
and
\be\widetilde
R^{\nu}_{\lambda\rho\sigma}=\partial_\rho\Gamma^\nu_{\lambda\sigma}-\partial_\sigma\Gamma^\nu_{\lambda\rho}
+\Gamma^\nu_{\tau\rho}\Gamma^\tau_{\lambda\sigma}-\Gamma^\nu_{\tau\sigma}\Gamma^\tau_{\lambda\rho}\,,
\label{2.10}\ee
while $i_\mu$  denotes the interior product which maps  $k$-forms
into $(k-1)$-forms. More exactly, if  $\alpha$ is the  $k$-form
\be\alpha=\frac{1}{k!}\alpha_{\mu_1\ldots\mu_k}dx^{\mu_1}\ldots
dx^{\mu_k}\,,\label{2.11}\ee
then
\be
i_\mu\alpha=\frac{1}{(k-1)!}\alpha_{\mu\mu_2\ldots\mu_k}dx^{\mu_2}\ldots
dx^{\mu_k}\,.\label{2.12}\ee

In order that \eqref{2.8} satisfies the properties of the Poisson
bracket, the following conditions must be imposed \cite{1}:\\
   a) $\th$  satisfies the Jacobi identity \eqref{2.3};\\
   b) $\widetilde\nabla_\rho$  is symplectic, i.e. $\widetilde\nabla_\rho\th=0$;\\
   c) the connection $\nabla_\nu$  has vanishing curvature, i.e. $[\nabla_\mu,\nabla_\nu]\alpha=0$  for any differential form
   $\alpha$;\\
   d) the curvature $\widetilde R^{\mu\nu}$ associated to the connection one-form  $\widetilde \Gamma^\rho_\sigma$ is covariantly  constant under
   $\nabla_\rho$, i.e.  $\nabla_\rho\widetilde R^{\mu\nu}=0$. The curvature $\widetilde R^{\mu\nu}$  is defined as in the Eqs. \eqref{2.9} -
\eqref{2.10}, but with $\Gamma^\rho_{\mu\nu}$ changed by
$\Gamma^\rho_{\nu\mu}$.

    Using the above properties, a star-product between differential forms has been defined in Ref. \cite{1}.
    Here we extend this definition to the case of Lie algebra valued differential forms. If  $\alpha=\alpha^aT_a$  and  $\beta=\beta^bT_b$ are two arbitrary
    such forms, where $\alpha^a$  and $\beta^b$  are ordinary differential forms of degrees  $ |\alpha|$ and respectively $ |\beta|$,
    then their star-product has the expression
\be\alpha\star\beta=\alpha\beta+\sum_{n=1}^\infty\left(\frac{i\hbar}{2}\right)^n
C_n(\alpha,\beta)
=\alpha^a\beta^b\,T_aT_b+\sum_{n=1}^\infty\left(\frac{i\hbar}{2}\right)^n\,C_n(\alpha^a,\beta^b)\,T_aT_b\,,\label{2.13}\ee
where $C_n(\alpha^a,\beta^b)$  are bilinear differential operators.
We impose then the condition that the star-product \eqref{2.13}
satisfies the property of associativity
\be       (\alpha\star\beta) \star\gamma= \alpha\star(\beta
\star\gamma)\,.\label{2.14}\ee
Introducing \eqref{2.13} in \eqref{2.14}, we find the following
general condition of associativity for an arbitrary order $n$:
\bea
&&C_n(C_0(\alpha,\beta),\gamma)+C_{n-1}(C_1(\alpha,\beta),\gamma)+C_{n-2}(C_2(\alpha,\beta),\gamma)\ldots+
C_0(C_n(\alpha,\beta),\gamma)\\
&&=C_n(\alpha,C_0(\beta,\gamma))+C_{n-1}(\alpha,C_1(\beta,\gamma))+C_{n-2}(\alpha,C_2(\beta,\gamma))\ldots+
C_0(\alpha,C_n(\beta,\gamma))\,. \nonumber\label{2.15}\eea
In Ref. \cite{1} the expressions of the operators
$C_n(\alpha^a,\beta^b)$ were obtained up to the second order in
$\theta$. We admit that these results are also valid in our case of
Lie algebra valued differential forms with adequate definitions.
They are
\bea
&&C_1(\alpha^a,\beta^b)\equiv\{\alpha^a,\beta^b\}=\th\left[\nabla_\mu\alpha^a\nabla_\nu\beta^b+(-1)^{
|\alpha|}\widetilde
R^{\rho\sigma}_{\mu\nu}(i_\rho\alpha^a)(i_\sigma\beta^b)\right],
\label{2.16}\\
&&C_2(\alpha^a,\beta^b)=\frac{1}{2}\th\theta^{\rho\sigma}\nabla_\mu\nabla_\rho\alpha^a\nabla_\nu\nabla_\sigma\beta^b
+\frac{1}{3}\theta^{\mu\rho}\partial_\rho\theta^{\nu\sigma}(\nabla_\mu\nabla_\nu\alpha^a\nabla_\sigma\beta^b-\nabla_\nu\alpha^a\nabla_\mu\nabla_\sigma\beta^b)\cr
&&-\frac{1}{2}\widetilde R^{\mu\nu}\widetilde R^{\rho\sigma}(i_\mu
i_\rho\alpha^a)(i_\nu i_\sigma\beta^b)-\frac{1}{3}\widetilde
R^{\mu\nu}(i_\nu \widetilde R^{\rho\sigma})[(-1)^{|\alpha|}(i_\mu
i_\rho\alpha^a)(i_\sigma\beta^b)+(i_\rho\alpha^a)(i_\mu
i_\sigma\beta^b)] \cr
&&+(-1)^{|\alpha|}\th\widetilde
R^{\rho\sigma}(i_\rho\nabla_\mu\alpha^a)(i_\sigma\nabla_\nu\beta^b)\,.\label{2.17}
\eea
It is important to observe that the operators
$C_n(\alpha^a,\beta^b)$ have the generalized Moyal symmetry
\cite{1},
\be
C_n(\alpha^a,\beta^b)=(-1)^{|\alpha||\beta|+n}C_n(\beta^b,\alpha^a)\,.\label{2.18}\ee
Taking into account the graded structure of our Poisson algebra, we
define the star commutator of two Lie algebra valued differential
forms $\alpha=\alpha^a\,T_a$ and $\beta=\beta^b\,T_b$ by
\be  [\alpha,\beta]_\star=\alpha\star\beta-
(-1)^{|\alpha||\beta|}\beta\star\alpha\, . \label{2.19}\ee
For example, if $\alpha$  and $\beta$  are Lie algebra valued
one-forms, we have
\be
 [\alpha,\beta]_\star=\alpha^a\beta^b[T_a,T_b]+\frac{i\hbar}{2}C_1(\alpha^a,\beta^b)\{T_a,T_b\}+\left(\frac{i\hbar}{2}\right)^2C_2(\alpha^a,\beta^b)[T_a,T_b]+\ldots  \label{2.20}\ee
This result shows that the star commutator of Lie-valued
differential forms does not close in general in the Lie algebra but
in its universal enveloping algebra. Exceptions are the unitary
groups whose Lie algebras coincide with their universal enveloping
algebras.
%However, we can extend the above results to any symmetry
%algebra considering the associated Hopf algebra. The universal
%enveloping of any Lie algebra can be always organized as a Hopf
%algebra.

We shall use all these properties in Sections 3 and 4 to develop a
noncommutative gauge theory with non-Abelian gauge group, up to the
second order in $\hbar$ (or equivalently ${\cal O}(\theta^3)$  ).

\section{Noncommutative gauge theory}

Consider the gauge group  $G$ whose infinitesimal generators satisfy
the algebra
 \be[T_a,T_b]=if^c_{ab}T_c,\ \ \ \ a,b,c=1,2,\ldots,m\,,\label{3.1}\ee
with the structure constants  $f^a_{bc}=-f^a_{cb}$ and the Lie
algebra valued infinitesimal parameter
\be\hat\lambda=\hat\lambda^aT_a.\label{3.2}\ee
We use the hat symbol to denote the noncommutative quantities of our
gauge theory. The parameter  $\hat\lambda$ is a 0-form, i.e.
$\hat\lambda^a$ are functions of the coordinates $x^\mu$  on the
symplectic manifold  $M$.

Now we define the gauge transformation of the noncommutative Lie
algebra valued gauge potential
\be\hat A=\hat A^a_\mu(x)\,T_a\,dx^\mu=\hat A_\mu\,dx^\mu,\ \ \ \hat
A_\mu=\hat A_\mu^a(x)\,T_a\,,\label{3.3}\ee
by
 \be\hat\delta\hat A= d\hat\lambda-i[\hat A,\hat\lambda]_\star\,.\label{3.4}\ee
Here we consider the formula \eqref{2.19} for the commutator  $[\hat
A,\hat\lambda]_\star$. Then, using the definition \eqref{2.13} of
the star-product, we can write \eqref{3.4} as
\be\hat\delta\hat A^a=d\hat\lambda^a+f^a_{bc}\hat
A^b\hat\lambda^c+\frac{\hbar}{2}d^a_{bc}C_1(\hat
A^b,\hat\lambda^c)-\frac{\hbar^2}{4}f^a_{bc}C_2(\hat
A^b,\hat\lambda^c)+\ldots\,,\label{3.5}\ee
where we denoted $\{T_a,T_b\}=d^c_{ab}T_c$. Since $\hat\lambda^a$
are functions, the operators $C_1(\hat A^b,\hat\lambda^c)$ and
$C_2(\hat A^b,\hat\lambda^c)$ have the expressions (see Eqs.
\eqref{2.16}-\eqref{2.17})
\bea &&C_1(\hat A^b,\hat\lambda^c)\equiv\{\hat
A^b,\hat\lambda^c\}=\th\nabla_\mu\hat
A^b\partial_\nu\hat\lambda^c\,,
\label{3.6}\\
&&C_2(\hat
A^b,\hat\lambda^c)=\frac{1}{2}\th\theta^{\rho\sigma}\nabla_\mu\nabla_\rho\hat
A^b\partial_\nu\partial_\sigma\hat\lambda^c+\frac{1}{3}\th\partial_\nu\theta^{\rho\sigma}(\nabla_\mu\nabla_\rho\hat
A^b\partial_\sigma\hat\lambda^c-\nabla_\rho\hat
A^b\partial_\mu\partial_\sigma\hat\lambda)\,.\nonumber\eea
Here we use the definition of the covariant derivative
\be \nabla_\mu\hat A^a=(\partial_\mu\hat
A^a_\nu-\Gamma^\rho_{\mu\nu}\hat
A^a_\rho)dx^\nu\equiv(\nabla_\mu\hat A^a_\nu)dx^\nu\,.\label{3.7}\ee
In particular, in the case when the gauge group $G$  is  $U(1)$, we
have  $f^a_{bc}=0$, $a,b,c=1$, $d^1_{11}=2$, $\hat A^1\equiv\hat A$,
$\hat\lambda^1\equiv\hat \lambda$, therefore \eqref{3.5} becomes
 \be\hat\delta\hat A=d\hat\lambda+\hbar C_1(\hat
A,\hat\lambda)+{\cal O}(\hbar^3)\,.\label{3.8}\ee
More explicitly, in terms of components,
\be\hat\delta\hat A_\mu=\partial_\mu\hat\lambda+\hbar
\theta^{\nu\sigma}\nabla_\nu \hat
A_\mu\partial_\sigma\hat\lambda+{\cal O}(\hbar^3)\,,\label{3.9}\ee
where $\hat A_\mu(x)$ is the $U(1)$  gauge potential and
$\hat\lambda(x)$ -- the infinitesimal parameter (phase). In zeroth
approximation \eqref{3.9} reproduces the usual $U(1)$ gauge
potential transformation.

Analogously, in the case of $U(2)$, we consider $a=(0,k)$,
$k=1,2,3$. Then, $f^k_{ij}=2\epsilon_{ijk}$,
$\{T_i,T_j\}=2\delta_{ij}T_0$, $i,j,k=1,2,3$ and
$\{T_0,T_k\}=\{T_k,T_0\}=2T_k$, $\{T_0,T_0\}=2T_0$, where  $T_0=I$
is the unit matrix and $T_k=\sigma_k$ -- the Pauli matrices as
generators of $SU(2)$. Then, since
\be\hat A=\hat A^k\sigma^k+\hat A^0 I,\ \ \
\hat\lambda=\hat\lambda^k\sigma_k+\hat\lambda^0 I\,,\label{3.10}\ee
we obtain from \eqref{3.5}
\bea \hat\delta\hat A^0&=&d\hat\lambda^0+\hbar\left[C_1(\hat
A^i,\hat\lambda^j)\delta_{ij}+C_1(\hat
A^0,\hat\lambda^0)\right]+{\cal
O}(\hbar^3)\,,\label{3.11}\\
\hat\delta\hat A^k&=&d\hat\lambda^k+2\epsilon_{ijk}\hat
A^i\hat\lambda^j+\hbar\left[C_1(\hat A^k,\hat \lambda^0)+C_1(\hat
A^0,\hat\lambda^k)\right]\cr
&-&\frac{\hbar^2}{2}C_2(\hat A^i,\hat A^j)\epsilon_{ijk}+{\cal
O}(\hbar^3)\,.\label{3.12}\eea
Here, the quantities $(\hat A^0,\hat\lambda^0)$  correspond to the
noncommutative $U(1)$  sector and  $(\hat A^k,\hat \lambda^k)$,
$k=1,2,3$ -- to the noncommutative  $SU(2)$ sector. Considering, in
addition, the restriction $U(2)\rightarrow U(1)$, we must keep only
Eq. \eqref{3.11}, but without the term $C_1(\hat
A^i,\hat\lambda^j)\delta_{ij}$, $i,j=1,2,3$. Thus, we rediscover the
previous $U(1)$ result \eqref{3.8} with $\hat A^0\equiv\hat A$,
$\hat\lambda^0\equiv\hat\lambda$.

We define the curvature two-form  $\hat F$ of the gauge potentials
by
\be\hat F=\frac{1}{2}dx^\mu dx^\nu\hat F_{\mu\nu}=d\hat
A-\frac{i}{2}[\hat A,\hat A]_\star\,.\label{3.13}\ee
Then, using the definition \eqref{2.13} of the star-product and the
property \eqref{2.18} of the operators  $C_n(\alpha^a,\beta^b)$, we
obtain from \eqref{3.13}
\be\hat F^a=d\hat A^a+\frac{1}{2}f^a_{bc}\hat A^b\hat
A^c+\frac{1}{2}\frac{\hbar}{2}d^a_{bc} C_1(\hat A^b,\hat
A^c)-\frac{1}{2}\frac{\hbar^2}{4}f^a_{bc} C_2(\hat A^b,\hat
A^c)+{\cal O}(\hbar^3)\,.\label{3.14}\ee
More explicitly, in terms of components we have
\be\hat F^a_{\mu\nu}=\partial_\mu\hat A^a_\nu-\partial_\nu\hat
A^a_\mu+f^a_{bc}\hat A^b_\mu\hat A^c_\nu+\frac{\hbar}{2}d^a_{bc}
C_1(\hat A^b_\mu,\hat A^c_\nu)-\frac{\hbar^2}{4}f^a_{bc} C_2(\hat
A^b_\mu,\hat A^c_\nu)+{\cal O}(\hbar^3)\,,\label{3.15}\ee
where we used the definition  $C_n(\hat A^b,\hat A^c)=C_n(\hat
A^b_\mu,\hat A^c_\nu)dx^\mu dx^\nu$, with
\bea C_1(\hat A^b_\mu,\hat
A^c_\nu)&=&\theta^{\rho\sigma}\left[\nabla_\rho\hat
A^b_\mu\nabla_\sigma \hat A^c_\nu-\widetilde
R^{\alpha}_{\sigma\mu\nu}\hat A^b_\rho\hat A^c_\alpha\right],
\label{3.16}\\
C_2(\hat A^b_\mu,\hat
A^c_\nu)&=&\theta^{\rho\sigma}\theta^{\lambda\tau}\Big[\frac{1}{2}\nabla_\rho\nabla_\lambda\hat
A^b_\mu\nabla_\sigma\nabla_\tau\hat
A^c_\nu+\frac{1}{3}(\nabla_\rho\nabla_\lambda\hat
A^b_\mu\nabla_\tau\hat A^c_\nu-\nabla_\lambda\hat
A^b_\mu\nabla_\rho\nabla_\tau\hat A^c_\nu)\cr
&-&\widetilde R^\alpha_{\tau\mu\nu}\nabla_\rho\hat
A^b_\lambda\nabla_\sigma\hat A^c_\alpha\Big]\,.\label{3.17} \eea

In the particular case of the $U(1)$  gauge group we obtain
\be\hat F_{\mu\nu}=\partial_\mu\hat A_\nu-\partial_\nu\hat
A_\mu+\hbar\theta^{\rho\sigma}\left[\nabla_\rho\hat
A^b_\mu\nabla_\sigma\hat A^c_\nu-\frac{1}{2}\widetilde
R^\alpha_{\sigma\mu\nu}\hat A^b_\rho \hat A^c_\alpha\right]+{\cal
O}(\hbar^3)\,.\label{3.18}\ee
Under the gauge transformation \eqref{3.4}, the curvature 2-form
$\hat F$ transforms as
\be\hat\delta\hat F=i[\hat \lambda,\hat F]_\star\,,\label{3.19}\ee
where we used the Leibniz rule
\be d(\hat \alpha\star\hat \beta)= d\hat
\alpha\star\beta+(-1)^{|\alpha|}\hat\alpha\star
d\hat\beta,\label{3.20}\ee
which we admit to be valid to all orders in  $\hbar$. In terms of
components, \eqref{3.19} reads
\be\hat\delta\hat F^a=f^a_{bc}\hat F^b\hat
\lambda^c+\frac{\hbar}{2}d^a_{bc} C_1(\hat
F^b,\hat\lambda^c)-\frac{\hbar^2}{4}f^a_{bc} C_2(\hat
F^b,\hat\lambda^c)+{\cal O}(\hbar^3)\,.\label{3.21}\ee
If the gauge group is  $U(1)$, then we obtain
$$
\hat\delta\hat F=\hbar C_1(\hat F,\hat
\lambda)=\hbar\theta^{\rho\sigma}\nabla_\rho\hat
F\partial_\sigma\hat\lambda+{\cal O}(\hbar^3)\,,
$$
or, in terms of components,
\be \hat\delta\hat F_{\mu\nu}=\hbar
\theta^{\rho\sigma}\nabla_\rho\hat F_{\mu\nu}\partial_\sigma\hat
\lambda+{\cal O}(\hbar^3)\,.         \label{3.22}\ee
Also, in the zeroth order, the formula \eqref{3.21} becomes
\be\delta F^a_{\mu\nu}=f^a_{bc} F^b_{\mu\nu}
\lambda^c\Longleftrightarrow\delta F=i[\lambda, F]\,.\label{3.23}\ee
This formula reproduces therefore the result of the commutative
gauge theory.
    Using again the Leibniz rule, we obtain the deformed Bianchi identity
\be d\hat F-i[\hat A,\hat F]_\star=0\,.\label{3.24}\ee
 If we apply the definition \eqref{2.19} of the star
commutator, we obtain
\be d\hat F+i[\hat F,\hat A]=\left[\frac{\hbar}{2}d^a_{bc} C_1(\hat
F^b,\hat A^c)-\frac{\hbar^2}{4}f^a_{bc}C_2(\hat F^b,\hat
A^c)\right]T_a+{\cal O}(\hbar^3)\,,\label{3.25}\ee
or, in terms of components,
\be d\hat F^a-fâ_{bc}\hat F^b,\hat A^c=\frac{\hbar}{2}d^a_{bc}
C_1(\hat F^b,\hat A^c)-\frac{\hbar^2}{4}f^a_{bc}C_2(\hat F^b,\hat
A^c)+{\cal O}(\hbar^3)\,.\label{3.26}\ee
We remark that in the zeroth order we obtain from \eqref{3.25} the
usual Bianchi identity
\be d F-i[ A, F]=0\,.\label{3.27}\ee
In addition, if the gauge group is $U(1)$, the Bianchi identity
\eqref{3.26} becomes
\be d \hat F=\hbar C_1(\hat A,\hat F)+{\cal
O}(\hbar^3)\,.\label{3.28}\ee
This result is also in accord with that of Ref. \cite{2}.

\section{Noncommutative Yang-Mills action}

Having established the previous results, we can construct a
noncommutative Yang-Mills action. Denote the metric in the
noncommutative space-time $M$ by  $G^{\nu\rho}$. Its covariant
derivative is
\be\nabla_\mu G^{\nu\rho}=\partial_\mu
G^{\nu\rho}+G^{\nu\sigma}\Gamma^{\rho}_{\mu\sigma}+\Gamma^\nu_{\mu\sigma}G^{\sigma^\rho}\,.\label{4.1}\ee
If $G^{\nu\rho}$ is not constant, we have to modify it to be a gauge
covariant metric  $\hat G^{\nu\rho}$ for the noncommutative
Yang-Mills action. The metric $\hat G^{\nu\rho}$  is gauge covariant
in the sense that it transforms like $\hat F$ (see \eqref{3.19})
\be\hat\delta\hat G^{\mu\nu}=i[\hat\lambda,\hat G^{\mu\nu}]_\star\,.
\label{4.2}\ee
Then, using the definition \eqref{2.19} of the star commutator, we
obtain from \eqref{4.2}
\be\hat\delta\hat G^{\mu\nu}=\hbar\theta^{\rho\sigma}\nabla_\rho\hat
G^{\mu\nu}\partial_\sigma\hat\lambda+{\cal O}(\hbar^3)\,.
\label{4.3}\ee
The explicit form of the metric   could be obtained using, for
example, the Seiberg-Witten \cite{4} map extended to the new
star-product defined in Ref. \cite{1} and used by us in developing
the noncommutative gauge theory.

Define the noncommutative Yang-Mills action by (see Ref. \cite{2})
\be\hat S_{NC}=-\frac{1}{2g^2}\langle Tr(\hat G\star\hat F\star \hat
G\star\hat F)\rangle=-\frac{1}{4g^2}\langle \hat
G^{\mu\rho}\star\hat F_{\rho\nu}\star\hat G^{\nu\sigma}\star\hat
F_{\sigma\mu}\rangle\,,\label{4.4}\ee
where $g$ is the gauge coupling constant, and we have used the
normalization property
\be Tr(T_aT_b)=\frac{1}{2}\delta_{ab}I\,.\label{4.5}\ee
Using the properties of gauge covariance \eqref{3.19} and
\eqref{4.3} for $\hat F$ and $\hat G$ respectively, we obtain
\be\hat\delta\hat S_{NC}=-\frac{1}{4g^2}\langle C_1(Tr(\hat G\hat
F\hat G\hat F),\hat\lambda)\rangle+{\cal
O}(\hbar^3)\,.\label{4.6}\ee
Now, since the integral is cyclic in the Poisson limit \cite{2},
i.e.
\be\langle C_1(Tr(\hat G\hat F\hat G\hat
F),\hat\lambda)\rangle=0\,,\label{4.7}\ee
Eq. \eqref{4.6} becomes
\be\hat\delta\hat S_{NC}=0\,.\label{4.8}\ee
Therefore, the action $\hat\delta\hat S_{NC}$  is invariant up to
the second order in $\theta$ (or $\hbar$). In Ref. \cite{2} it has
been proven that the action \eqref{4.4} can be further simplified as
\be\hat S_{NC}=-\frac{1}{2g^2}\langle Tr(\hat G\hat F\hat G\hat
F)\rangle+{\cal O}(\hbar^3)\,.\label{4.9}\ee
Imposing then the variational principle  $\hat\delta_{\hat A} \hat
S_{NC}=0$ with respect to the noncommutative gauge fields  $\hat
A^a_\mu$, we can obtain the noncommutative Yang-Mills field
equations.

\section{ Discussion}

We have developed a noncommutative gauge theory by using a
star-product between differential forms on symplectic manifolds
defined as in Ref. \cite{1}, by extending the definition given in
Ref. \cite{1} to the case of Lie algebra valued differential forms.
In this manner we have obtained a graded Lie algebra valued Poisson
algebra where the star-bracket operation can be both commutator and
anti-commutator, depending on the grades of the two forms. The
graded, but not Lie algebra valued, Poisson algebra on a symplectic
manifold was initially introduced in Refs. \cite{2} and \cite{3}. In
these papers an explicit form of the bracket for one-forms has been
obtained. In Ref. \cite{1} the results have been generalized to the
case of graded Poisson bracket for arbitrary degrees of differential
forms. In order to develop a noncommutative gauge theory we have
defined the star commutator of two Lie algebra valued differential
forms. Since the star-product does not close in general in the Lie
algebra, but only in its universal enveloping algebra, we can use
the unitary Lie algebras $U(n)$ as gauge symmetry or extend the
results to the Hopf algebra for any other algebras. We have
introduced the noncommutative one-form gauge potentials $\hat A$ and
the field strength two-form $\hat F$ and have obtained their gauge
transformation laws. We have proven that the defined field strength
$\hat F$ is gauge covariant and satisfies a deformed Bianchi
identity. To obtain these results we have used the Leibniz rule
which we admit to be valid to all orders in $\theta$.

Finally, we have defined an action for the gauge fields by
introducing a gauge covariant noncommutative metric $\hat
G^{\nu\rho}$  on the space-time manifold. The gauge invariance of
this action has been verified up to the second order in  $\theta$
using the property that the integral is cyclic in the Poisson limit
\cite{2}. The explicit form of the metric $\hat G^{\nu\rho}$ could
be obtained using, for instance, the Seiberg-Witten map extended to
the new star-product. Extending the gauge theory to higher orders in
$\theta$ requires to find the explicit expressions for the bilinear
differential operators $C_n(\alpha^a,\beta^b)$ which define the
star-product.

\section*{Acknowledgements}
We would like to thank Markku Oksanen for useful comments. The
support of the Academy of Finland under the Projects No. 121720 and
127626 is acknowledged. G. Z. acknowledges the support of
CNCSIS-UEFISCSU Grant No. 620 of the Ministry of Education and
Research of Romania.

\end{document}